\documentclass[review]{elsarticle}

\usepackage{hyperref}
\usepackage{makecell}

\begin{document}

\begin{frontmatter}

\title{Imaging and time stamping of photons with nanosecond resolution in Timepix based optical cameras}

\author{Andrei Nomerotski}
\address{Brookhaven National Laboratory, Upton NY 11973, USA}




\begin{abstract}
This contribution describes fast time-stamping cameras sensitive to optical photons and their applications.
\end{abstract}


\end{frontmatter}


\section{Imaging with photon counting}

Imaging of fast processes with nanosecond-scale timing resolution is necessary in many applications. Detection of individual photons is ultimately the best route to capture all available information for the process in which they were created. This information then can be used to do basic imaging by counting all detected photons or, if desired, to perform more complex operations with the data, which could involve timing of individual hits or some correlation analysis. Photon counting is already a widely used modality in x-ray imaging, where the signal is large enough to detect individual photons directly without external amplification, and to enable measurement of the time and energy for each of them. Here, we describe the concept of time stamping for optical photons, including single optical photons, and review recent results on TimepixCam and Tpx3Cam cameras, which can be used for this purpose. 

{\bf Framing versus time-stamping}: In conventional imagers the signal is integrated in a slice of time and stored in the pixel for consequent readout, frame by frame. Currently, this approach makes it impossible to achieve nanosecond resolution for continuous readout because the data rates are becoming prohibitively high. 
Silicon based imagers for the 400-1000nm wavelength range are commercially available for frame rates of up to 10kHz, with their pixel output rate of Gpix/s. This corresponds to the time resolution of 100 $\mu$s, which is orders of magnitude inferior to the nanosecond-scale requirement.  More specialized cameras based on CCD or CMOS technology are becoming available, which can achieve faster rates by buffering multiple frames in the sensor \cite{etoh2013, kirana}. However, this approach is not scalable and cannot be used for continuous readout. When the buffer is full one must stop and transfer the data to the outside world so the overall duty cycle is small.

An alternative approach is based on the so-called data-driven readout when only the data of interest is captured. To reduce the rate and support continuous readout, only those pixels in which the signal exceeds a certain predefined level are measured and read out. Of course, this approach is favored over the full frame readout  when the occupancy of the sensor is not too high, typically less than 10-20\%. In the context of light detection, this technique was implemented in the PImMS camera for mass-spectrometry applications \cite{pimms1, pimms2}, and, as described in the following, in the optical cameras based on readout chips of the Timepix family \cite{timepix, timepix3}. 

In the data-driven approach the signal shaping in the front-end electronics is fast, with peaking time of $\sim100$ ns in order to be compatible with nanosecond timing resolution. With enough statistics, images can be formed by counting the photons, and also more complex analyses can be performed. For example, one could use the measured information  to determine time coincidences between the photons, to calculate correlations, invariant masses etc. This has close parallels with the registration of x-rays at the synchrotron light sources and of ionizing particles in the high energy physics experiments, where they are detected as standalone objects. 

The high rate capability of the readout electronics is essential for fast accumulation of statistics. Assuming a similar back-end bandwidth as for the framing approach, Gpix/sec, and a 0.1-1 Mpix array, it is easy to calculate that this would correspond to an average pixel and, therefore, an average frame readout rate of 1-10 kHz. Thus, the average "frame" rate for a data-driven system is similar to the discussed above framing approach but in addition to the "normal" imaging one has a precise time-stamp for each photon. The price for this, of course, is complexity of the pixel as the data-driven approach needs to accommodate considerably more than just a few transistors as in simple framing architectures. It is typical to have hundreds of transistors per pixel. This also leads to a larger pixel size: 55x55 and 70x70 square micron respectively for Timepix and PImMS sensors. An overview comparison of these time-stamping optical cameras is given in Table \ref{tab:comparison}, with more detail on the Timepix-based cameras provided in the further sections.

\begin{table}[h!]
  \begin{center}
    \label{tab:comparison}
    \begin{tabular}{|c|c|c|c|c|} 
      \hline
        			  			& PImMS-1		& PImMS-2	& TimepixCam	& Tpx3Cam\\
      \hline
	year					& 2009			& 2012		& 2015		& 2017 		 \\
	array					& 72~x~72		& 324~x~324	& 256~x~256	& 256~x~256\\
	pixel size, $\mu$m$^2$	& 70~x~70		& 70~x~70	& 55~x~55	& 55~x~55	\\
	time resolution			& 12.5~ns			& 12.5~ns		& 10~ns		& 1.6	~ns		\\
	pixel information		& TOA			& TOA		& TOA		& TOA \& TOT  \\
	max rate				& 1200~Hz		& 60~Hz		& 2000~Hz	& 80~Mpix/s	\\
	max QE, \%	& 8			& 8			& 90			& 90 			\\
	technology			& MAPS \cite{pimms2}		& MAPS 		& hybrid CMOS & hybrid CMOS \\
      \hline
    \end{tabular}
   \caption{Comparison of the existing time-stamping optical cameras.}
   \label{tab:milestones}
  \end{center}
\end{table}

\begin{figure*}[!htb]
	\centering
	\includegraphics[width=0.8\linewidth]{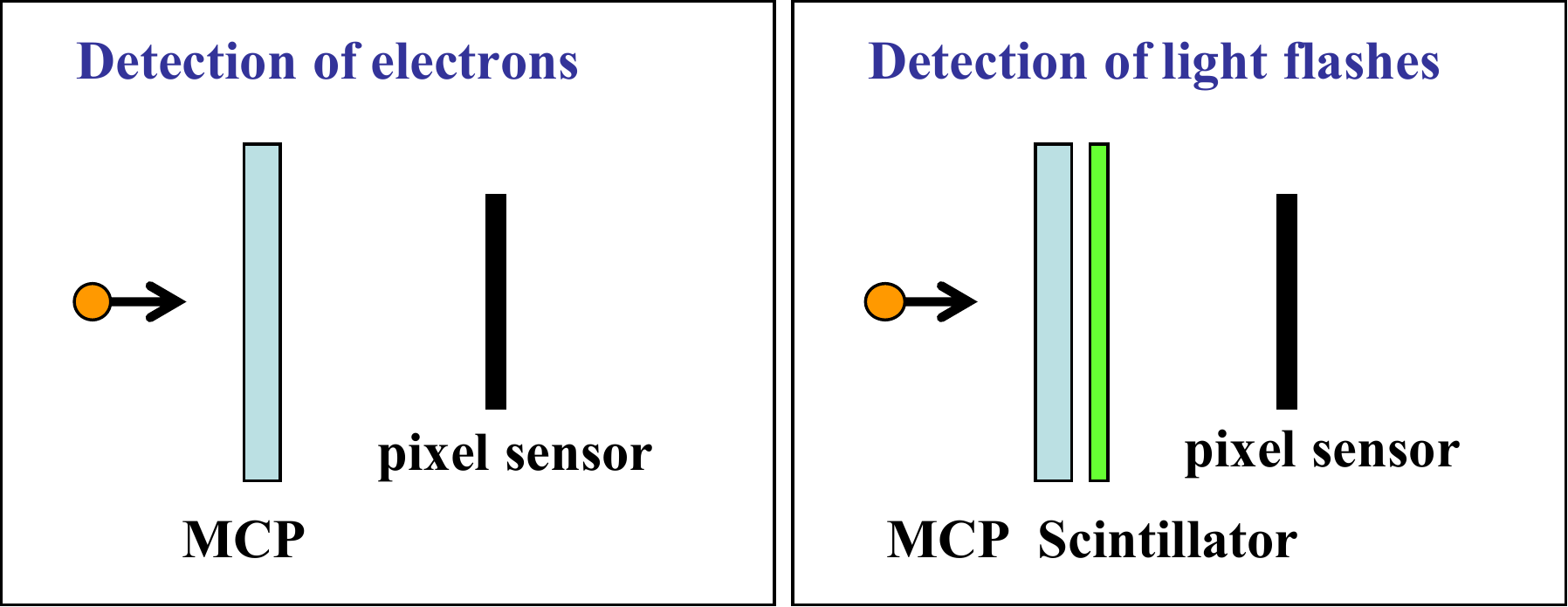}
	\caption{\footnotesize{Comparison of direct and optical detection.}}
	\label{fig:optical}
\end{figure*}

{\bf Optical versus direct detection}:  The pixel noise is too high to be sensitive to single particles. Their detection for the applications discussed in this paper currently require external amplification in the form of a micro-channel plate (MCP), which produces an avalanche of electrons in the MCP pores. In principle, these electrons can be collected to sense electrodes on the readout chip directly. Indeed, the Timepix2 chip was used before to directly detect electrons from the MCP both for optical photons and for ion imaging applications \cite{Jungmann2010, Jungmann2013, valerga2014}. Another approach advocated here would be to send the MCP electrons to a thin layer of fast scintillator so they produce a flash of light, which can be registered with an optical camera. This approach is very common in the ion imaging and in the intensified cameras for night vision, though it is normally used for much slower scintillators and cameras. The two approaches are illustrated in Figure \ref{fig:optical}.

Both PImMS and Timepix cameras adopted the optical approach, which has three important advantages over the direct detection approach. Firstly, the direct collection of MCP electrons requires  close ($\sim$ mm) proximity of the readout chip, which must be placed inside the experimental volume, to the kV-scale voltages at the MCP. This is a considerable complication because of possible sparking. It also requires a solution for cooling of the chip in a vacuum. The optical approach completely avoids these issues. Secondly, the camera is placed outside of the vacuum, so it is fully decoupled from the rest of the setup. In many cases it just replaces a slower camera used in these experiments before, and furthermore, the approach allows for a straightforward upgrade path as technology improves, without interfering with the rest of the experiment. Lastly, the optical approach allows flexible mapping between the scintillator screen and sensor by introducing demagnification, so a larger scintillator can be fully imaged in a small sensor. Other optical schemes with magnification, relay lenses and mirrors are also possible.

\section{Fast optical cameras based on Timepix2 and Timepix3}

\begin{figure*}[!htb]
	\centering
	\includegraphics[width=0.50\linewidth]{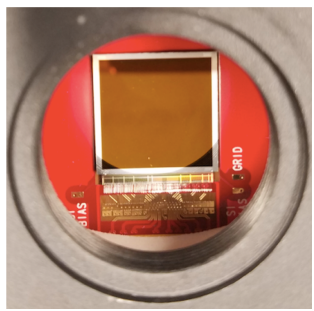}
	\includegraphics[width=0.50\linewidth, angle =270]{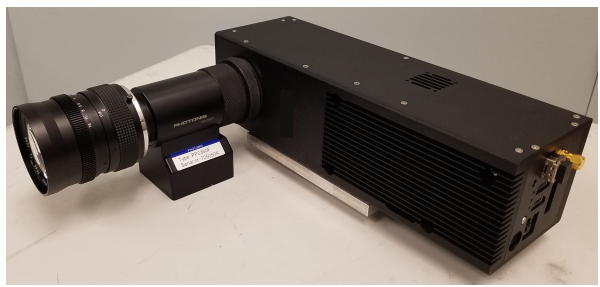}	
	\caption{\footnotesize{Top: optical sensor inside Tpx3Cam. Bottom: intensified version of the camera with attached image intensifier and 50 mm Navitar lens.}}
	\label{fig:camera}
\end{figure*}

The fast cameras described below are based on the so-called hybrid pixel detectors: a pixelated optical sensor with a high quantum efficiency (QE) is bump-bonded to a Timepix ASIC (application specific integrated circuit). The design of the back-side illuminated silicon sensor, in particular its thin entrance window, was inspired by the fully depleted astronomical CCDs, such as used, for example, in LSST \cite{Radeka2009}, while the readout chip is a product of the Medipix Collaborations\footnote{The Timepix2 chip used in TimepixCam was developed by the Medipix2 collaboration, while Timepix3 used in Tpx3Cam was developed by the Medipix3 collaboration.} \cite{medipix} led by CERN employing technologies developed for the LHC experiments. The data acquisition system for the cameras used commercial readouts of x-ray detectors. The first fast camera based on the Timepix2 readout chip \cite{timepix}, TimepixCam \cite{timepixcam}, was built in 2015, followed in 2017 by the next generation camera, Tpx3Cam \cite{tpx3cam}, based on Timepix3 \cite{timepix3}. Both cameras employ the same sensor. Figure \ref{fig:camera} shows the optical sensor inside Tpx3Cam and intensified version of the camera with attached image intensifier and 50 mm f/0.95 Navitar lens. Below we provide more details about Tpx3Cam as the most advanced camera. 

In Timepix3 each pixel has a predefined threshold so only the pixels above the threshold perform the time measurements and are read out, making the chip completely data-driven. It uses a free running 40MHz clock  and does not require any external triggering. The fired pixels provide the time-of-arrival (TOA) information with 1.56 ns granularity, and time-over-threshold (TOT) information with 25 ns granularity. The individual pixel deadtime is equal to the pixel TOT + 475 ns.

\begin{figure*}[!htb]
	\centering
	\includegraphics[width=0.8\linewidth]{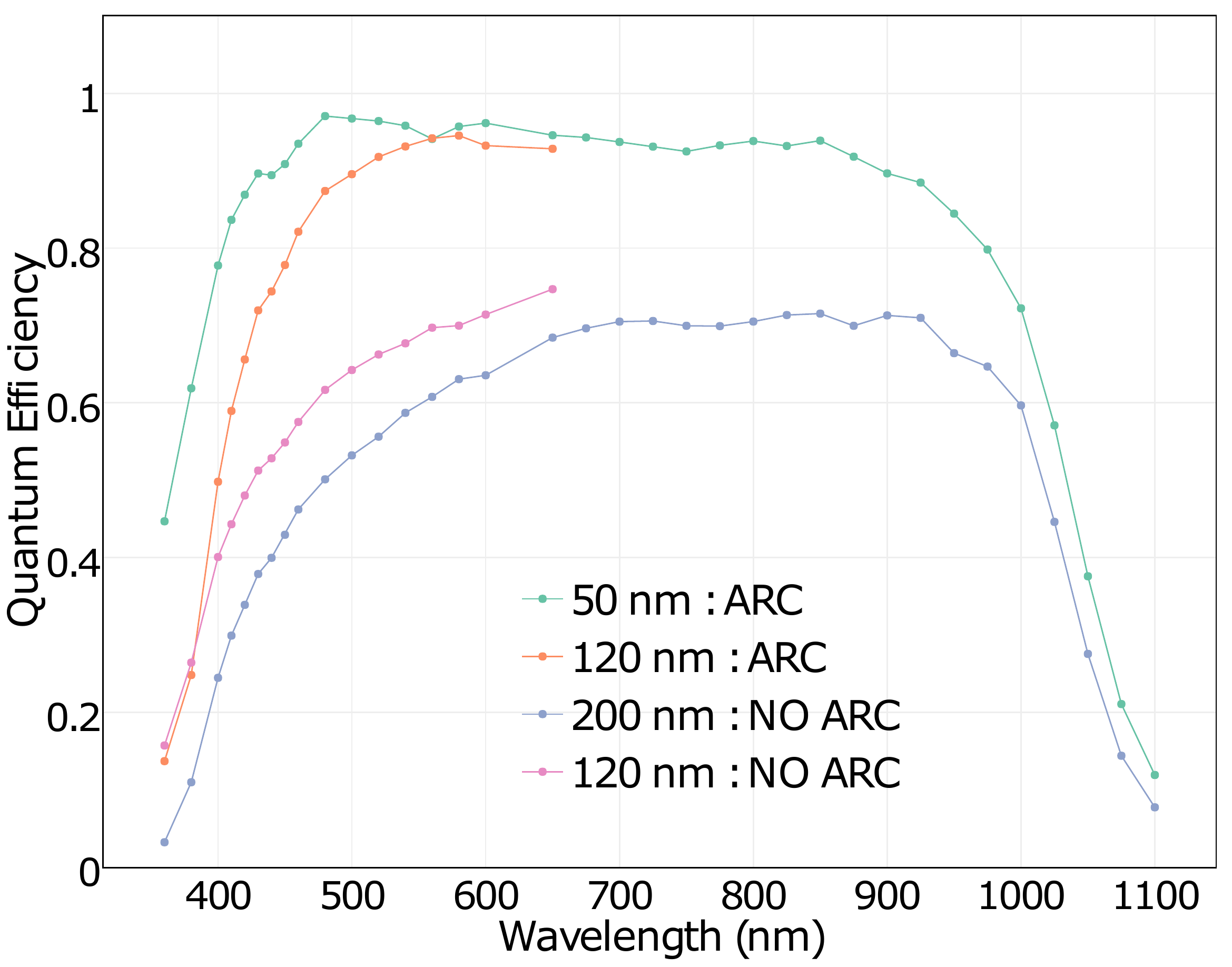}
	\caption{\footnotesize{Quantum efficiency of the Timepix compatible optical sensor as a function of wavelength for several types of sensors with varying thickness of passivation layer and with/without anti-reflective coating (ARC).}}
	\label{fig:QE}
\end{figure*}

Fast scintillator P47, used in combination with MCP has the rise and decay time of 7ns and 100ns, respectively, and maximum emission in blue at 430 nm \cite{P47}. The absorption depth for the blue photons is only 250 nm, so it is very important that the passive layer on the sensor surface is thin enough to let the photons through. At the same time it should be conductive to ensure uniform electric field and full depletion of the 300 micron thick silicon sensor because the created charge carriers need to drift from the sensor window on the back side to the collection pads on the front-side, which are bump-bonded to Timepix.  In this case holes are collected as the sensor is of a p-on-n type.  Anti-reflective coating on the sensor was optimized for the P47 emission spectrum. Figure \ref{fig:QE} shows the measured quantum efficiency as a function of the wavelength for several types of sensors with varying thickness of passivation layer and with/without  anti-reflective coating (ARC). The measurements are described in detail in \cite{characterizationTpxCam} together with other test results for the optical sensors.

The Tpx3Cam camera readout is based on the SPIDR data acquisition system \cite{spidr1, spidr2}, which is available commercially \cite{ASI}. The SPIDR maximum output rate is 80 Mpix/s. It also implements a time-digital-converter (TDC), which provides a time-stamp with 0.26 ns granularity for an input signal and is synchronized to the Timepix3 data. This input can provide a precise time reference for the Timepix3 hits using, for example, a pulse synchronous to the laser employed in the experiments. It also can be used to synchronize multiple cameras to each other and to external devices.

\subsection{Detection of single photons}

\begin{figure*}[!htb]
	\centering
	\includegraphics[width=0.8\linewidth]{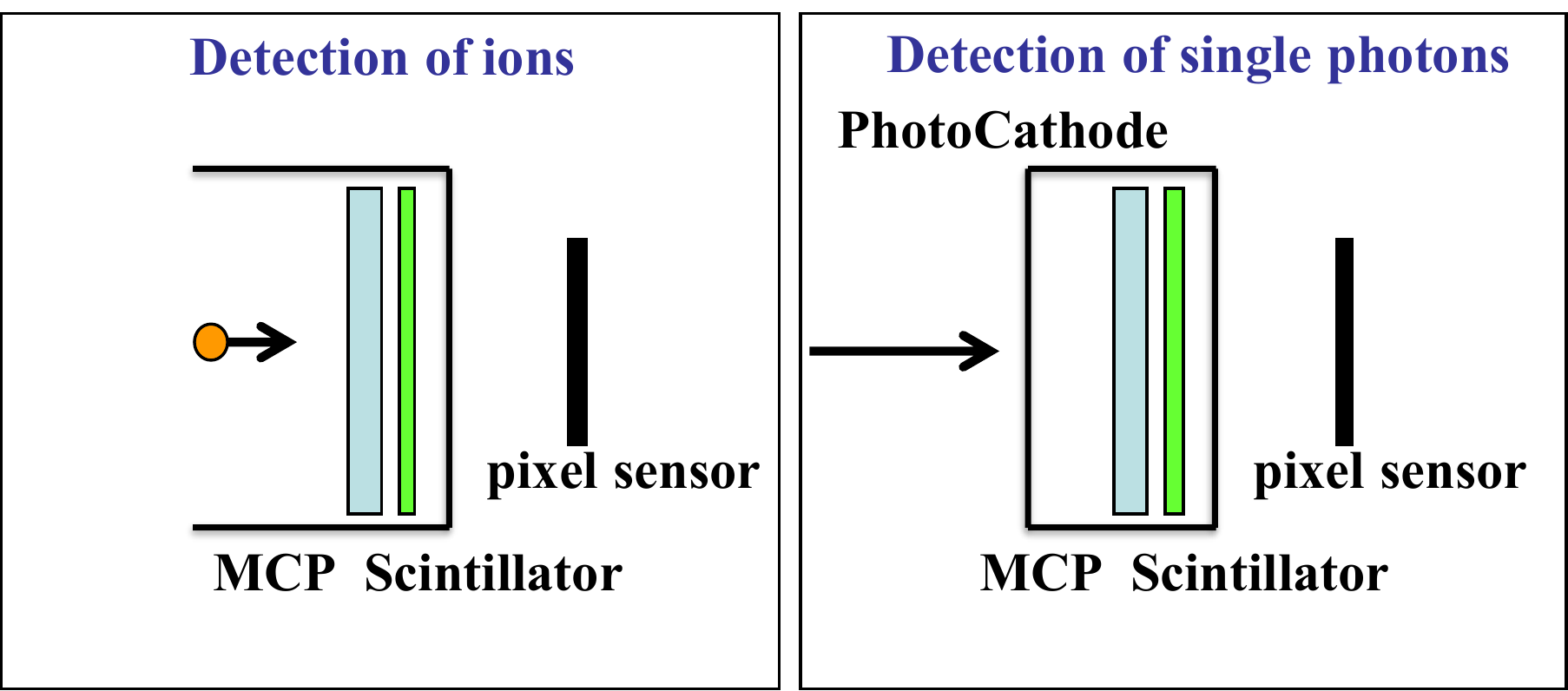}
	\caption{\footnotesize{Comparison of ion and single photon detection using the MCP and fast scintillator.}}
	\label{fig:single}
\end{figure*}

Detection of single photons requires external amplification, so the intensified version of the camera employs an image intensifier, an off-the-shelf vacuum device with photocathode followed by a MCP and  scintillator. The MCP can operate at gains up to few times 10$^5$ making the device sensitive to single photons. The back-end of the intensifier used for the measurements discussed below consists of the MCP/P47 assembly, so it is very similar to the assembly routinely used for the ion imaging, as illustrated in Figure \ref{fig:single}. This similarity in detection of various particles: ions, electrons, single photons and, as proposed below, x-rays and neutrons, makes this approach very versatile, because the same camera can be used for all of them without any modifications. 

\begin{figure*}[!htb]
	\centering
	\includegraphics[width=0.9\linewidth]{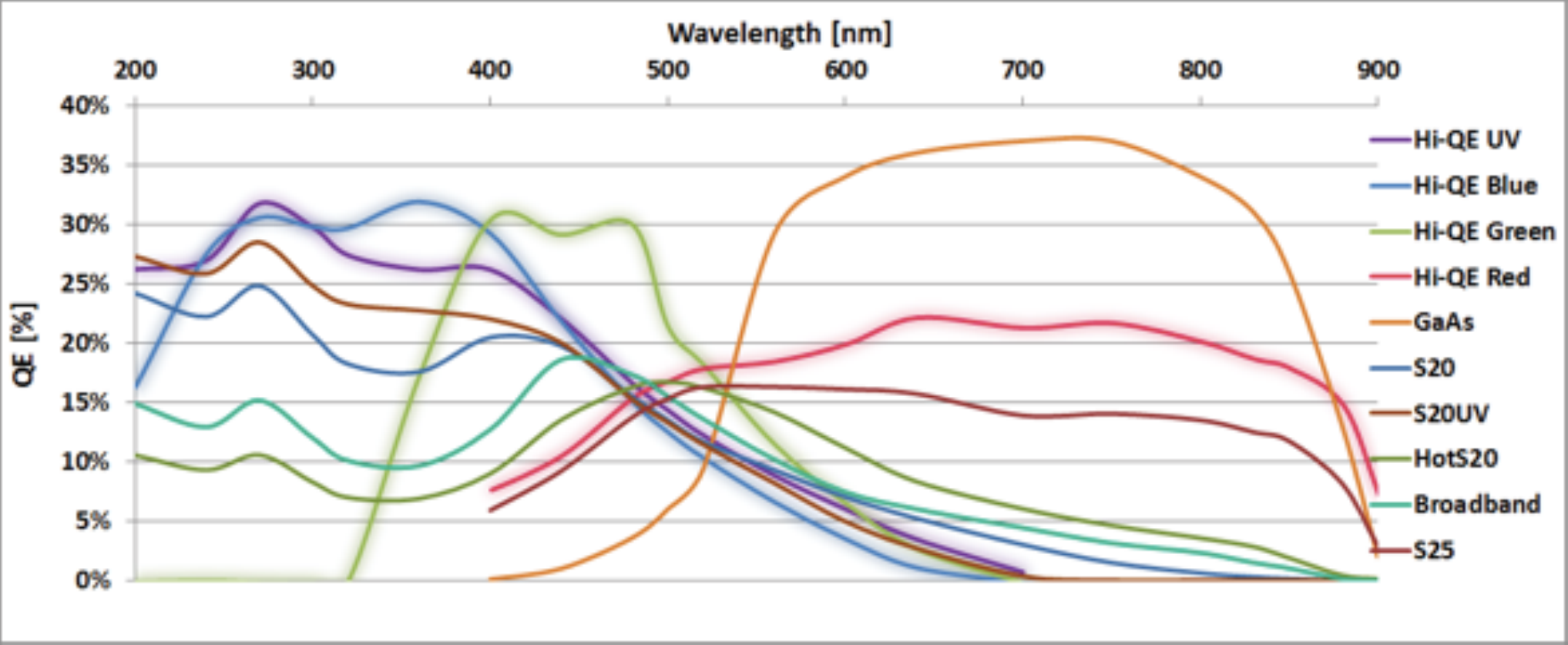}
	\caption{\footnotesize{Quantum efficiency for  photocathodes available from Photonis, as a function of wavelength.}}
	\label{fig:photonisQE}
\end{figure*}

The  photocathode in the intensifier can be selected to match the QE requirements for a particular application. There is a wide choice of available photocathodes with different spectral sensitivities. An example is given in Figure \ref{fig:photonisQE}, which provides QE as function of wavelength for a variety of Photonis photocathodes \cite{Photonis}. The 18mm Photonis intensifier tested with the camera had High-QE Red photocathode with QE equal to $18\%$ at 800 nm, the wavelength relevant for the quantum information science applications \cite{qis2018}, and dark count rate of $\sim80$ kHz over the full photocathode area at room temperature.

The intensifier in a cricket \cite{cricket} is shown together with the camera in the right part of Figure \ref{fig:camera}. The cricket integrates into a single unit the intensifier, power supply and collimating relay optics between the intensifier output and camera sensor. The cricket is fetched with C-mounts on both ends for attachment to the camera and to a lens. 

\begin{figure*}[!htb]
	\centering
	\includegraphics[width=0.49\linewidth]{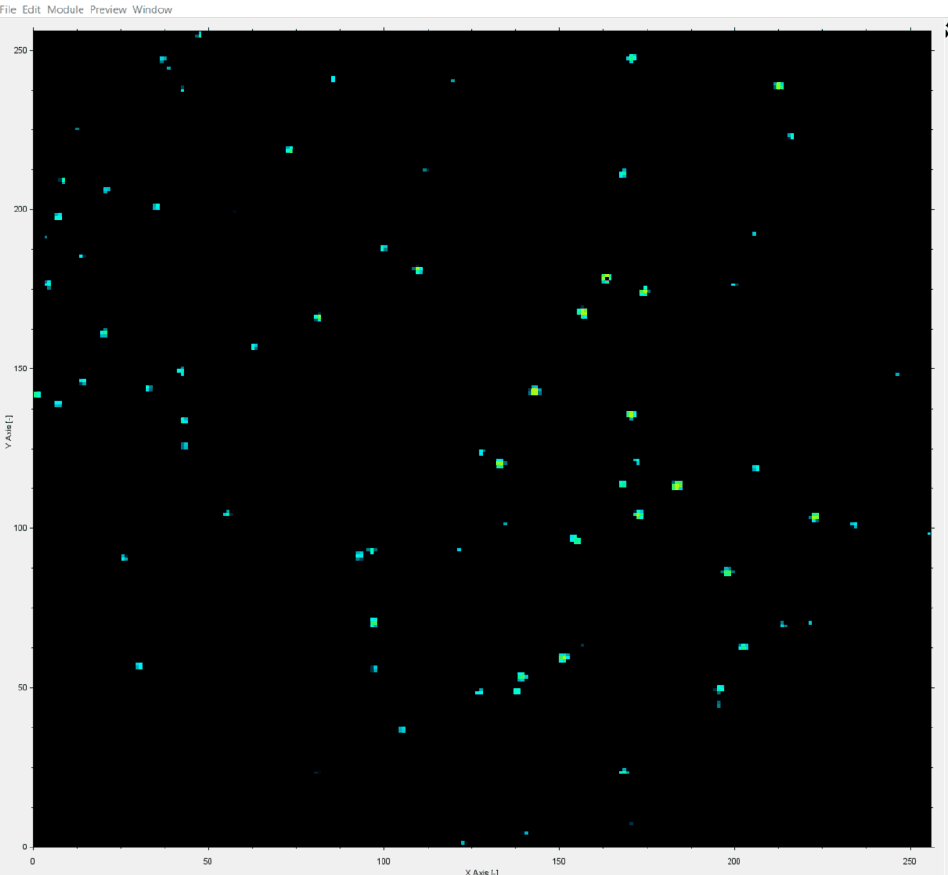}
	\includegraphics[width=0.49\linewidth]{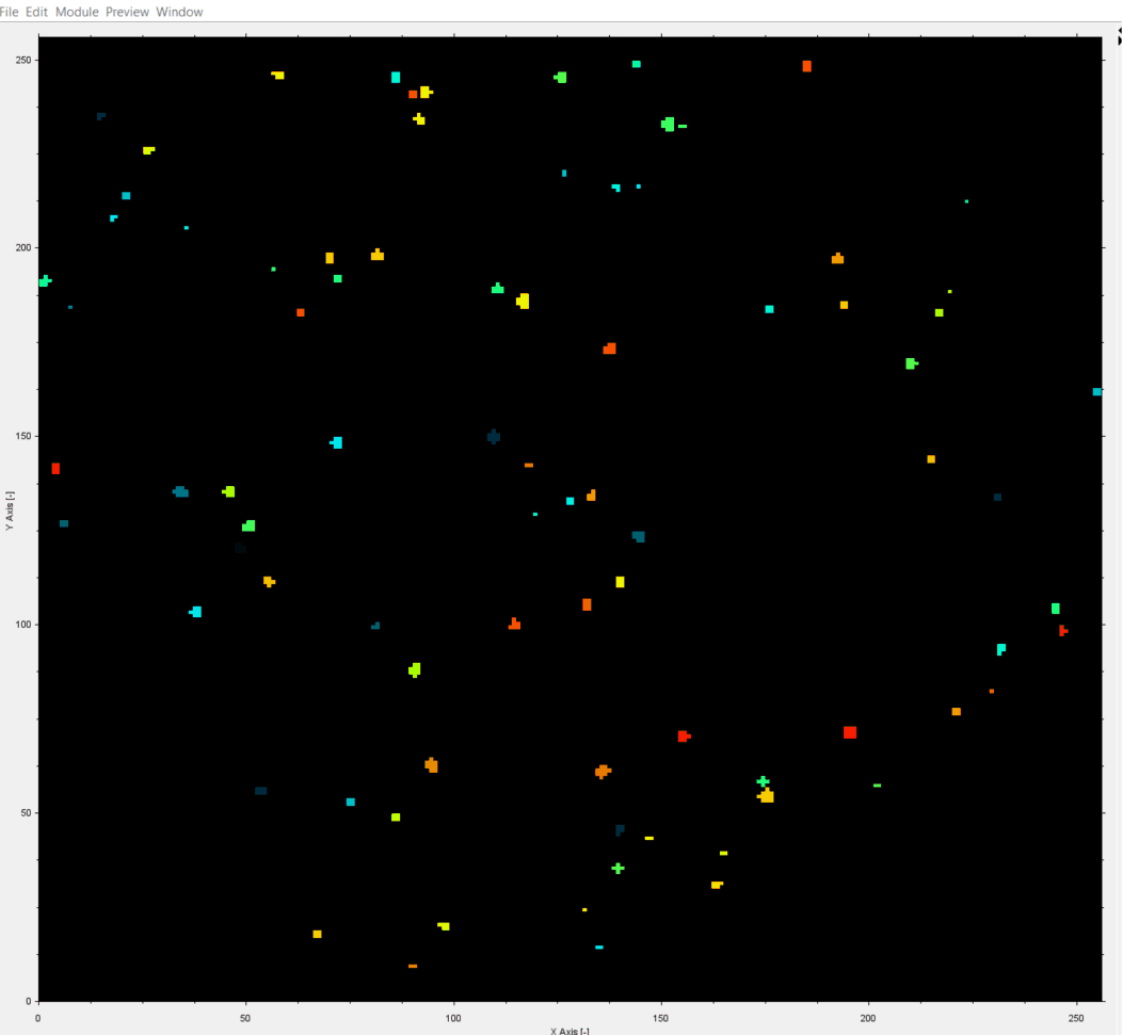}
	\caption{\footnotesize{Examples of single photon hits recorded by the camera in a time slice of $5~\mathrm{ms}$. The hits are shown  as heatmaps in TOT representation (left) and TOA representation (right).}}
	\label{fig:photons1}
\end{figure*}

Figure \ref{fig:photons1} shows examples of single photon hits recorded by the camera in a time slice of $5~\mathrm{ms}$. The hits are shown  as heatmaps in TOT representation (left) and TOA representation (right). One can see that each hit typically consists of several pixels, which have signals above the threshold. Since all fired pixels measure TOA and TOT independently and have position information, it can be used for centroiding to determine the photon coordinates. The centroiding considerably improves the spatial resolution, easily to sub-pixel values. The timing resolution can be improved at the pixel level correcting for the time-walk, an effect caused by the dependence of the front-end pixel electronics time response on the amplitude of the input signal \cite{Turecek_2016, tpx3cam}. Since the latter is measured as TOT, the TOA can be corrected achieving 2 ns timing resolution (rms) \cite{tpx3cam, qis2018}. The resolution per photon can be further improved by combining timing information in multiple pixels, which belong to the same photon hit.

\subsection{Applications}

During the last three years the Timepix-based time-stamping cameras have been used in a variety of different applications, briefly reviewed below.

{\bf Ion and electron imaging}: In this case the ions impinge a micro-channel plate producing light flashes in the fast scintillator behind the MCP, which are imaged by the Timepix camera. This approach works particularly well for the coincidence velocity map imaging (VMI), an essential tool in the study of reaction dynamics and strong field laser-matter interactions. VMI projects the transverse momenta of charged particles to positions on a 2D detector such that for a given particle species, its distance to the center of the detector is proportional to the initial transverse velocity. It also requires simultaneous measurement of all reaction fragments, which is not possible with a slow camera. The measurements with TimepixCam and Tpx3Cam were performed at FLASH in DESY \cite{flash2018} and in the ultrafast spectroscopy lab at Stony Brook University \cite{tpx3cam}. Several other groups interested in the ion and electron imaging used the cameras with the data analysis currently in progress.

{\bf Phosphorecent lifetime imaging}: The camera performance for single photons was first validated in experiments on the photon counting phosphorescence lifetime imaging. In these types of measurements the camera is time-stamping photons, which are emitted with characteristic lifetime after excitation of a material under study with a short laser pulse. The results are described in \cite{FLIM1} and \cite{FLIMCFN}.

{\bf Imaging of entangled photons}: in the single photon mode the camera was also used for characterization of sources of entangled photons and for measurements of their temporal and spatial correlations \cite{qis2018}.

{\bf Optical readout of TPC}: recently Tpx3Cam was employed for the first demonstration of three-dimensional optical readout of a dual phase liquid Ar TPC (time projection chamber) \cite{tpc2018}. The approach is described in detail in \cite{tpc2009, tpc2015}.

{\bf Other possible applications}: the camera can be used to register light flashes in thin scintillators produced by x-rays with energy 10keV and higher, where the direct detection with silicon sensors becomes inefficient. This approach, currently under testing,  requires an intensified version of the camera since only a handful of photons will be collected per x-ray. However, it should allow a simple technique of time-stamping for individual x-rays with nanosecond resolution. Also it will avoid placing the detector in the direct beam of x-rays. A similar approach of registration in a thin scintillator can be employed for neutrons.

\section{Future R\&D directions}

It is not possible to detect a single optical photon with good time resolution in a silicon sensor without amplification because of the  noise. The amplification can be achieved outside of the sensor using an image intensifier, as described above or, alternatively, inside the sensor. Sensors with internal amplification are based on technologies such as SPAD (single photon avalanche devices) \cite{Perenzoni2016} and LGAD (low gain avalanche devices)  \cite{Pellegrini2016, lgadbnl}, which have a multitude of  applications, including high-energy physics. Advances in the CMOS technology are enabling integration of these devices in to pixelated silicon sensors. This lead to production of the first SPAD arrays capable to count and time stamp single photons with resolution below 100 ps \cite{Lee2018}. Infrared HgCdTe imagers with internal amplification, which are compatible with single photon detection, also  received a lot of attention in astronomy in the context of extremely low light level applications \cite{saphira}. In terms of  charge collection those sensors should be able to support fast timing. All the above sensors should be also compatible with the hybrid approach as presented here, which will benefit from future improvements in readout ASICs as, for example, in Timepix4, which is aiming at the 200 ps timing resolution \cite{timepix4}.

\section{Acknowledgements}

The author is grateful to all his collaborators who helped in the implementation and testing of the first cameras, in particular, to Merlin Fisher-Levine, Peter Svihra, Martin van Beuzekom, Irakli Chakaberia, Peter Takacs, Thomas Tsang, Bram Bouwens, Erik Maddox, Jan Visser, Arthur Zhao, Thomas Weinacht,  Vaclav Vrba, Zdenko Janoska, Mael Flament, Eden Figueroa, Daniel Rolles, Rebecca Boll, Heinz Graafsma, David Pennicard, Milija Sarajlic, Jochen Kupper, Igor Rubinski, Sebastian Trippel, Ruth Livingstone, Liisa Hirvonen, Klaus Suhling, Adam Roberts, Kostas Mavrokoridis,  Mark Brouard, Michael Burt and Ilia Vardishvili. This work was supported by the BNL LDRD grants 13-006 and 18-030.

\section*{References}

\bibliography{ulitima_full}

\end{document}